\documentclass[preprint]{aastex}

\shorttitle{Variable stars in the open cluster M11}
\shortauthors{Koo et al.}

\begin{document}
\title{Variable stars in the Open Cluster M11 (NGC 6705)}
\author{ J.-R. Koo}
\affil{Department of Astronomy and Space Science, Chungnam National University, 
\\Daejeon 305-764, Korea; and Korea Astronomy and Space Science Institute, 
\\Daejeon 305-348, Korea}
\email{koojr@cnu.ac.kr}
\author{ S.-L. Kim}
\affil{Korea Astronomy and Space Science Institute, Daejeon 305-348, Korea}
\email{slkim@kasi.re.kr}
\author{ S.-C. Rey}
\affil{Department of Astronomy and Space Science, Chungnam National University, \\Daejeon 305-764, Korea}
\author{ C.-U. Lee}
\affil{Korea Astronomy and Space Science Institute, Daejeon 305-348, Korea}
\author{ Y. H. Kim}
\affil{Department of Astronomy and Space Science, Chungnam National University, \\Daejeon 305-764, Korea}
\author{ Y. B. Kang}
\affil{Department of Astronomy and Space Science, Chungnam National University, 
\\Daejeon 305-764, Korea; and Korea Astronomy and Space Science Institute, 
\\Daejeon 305-348, Korea}
\and
\author{ Y.-B. Jeon}
\affil{Korea Astronomy and Space Science Institute, Daejeon 305-348, Korea}

\begin{abstract}
$V$-band time-series CCD photometric observations of the intermediate-age open cluster M11 were performed 
to search for variable stars. Using these time-series data, we carefully examined 
light variations of all stars in the observing field. A total of 82 variable stars were discovered, 
of which 39 stars had been detected recently by \citet{hargis2005}. 
On the basis of observational properties such as variable period, light curve shape, and position on a
color-magnitude diagram, we classified their variable types as 11 $\delta$ Scuti--type pulsating stars,
2 $\gamma$ Doradus--type pulsating stars, 40 W UMa--type contact eclipsing binaries, 13 Algol--type
detached eclipsing binaries, and 16 eclipsing binaries with long period.
Cluster membership for each variable star was deduced from the previous proper motion results 
\citep{mcnamara1977} and position on the color-magnitude diagram. 
Many pulsating stars and eclipsing binaries in the region of M11 are probable members of the cluster.
\end{abstract}


\section{Introduction}
A number of pulsating variable stars are found in stellar clusters. Population I variables such as $\delta$ Scuti--type stars and 
$\beta$ Cephei--type stars are populous in open clusters and Population II variables such as RR Lyrae--type stars and 
SX Phoenicis--type stars are rich in globular clusters. Pulsating stars in clusters are important targets for investigating 
stellar internal structure and evolution (so-called asteroseismology) because physical parameters such as age and absolute magnitude 
which impose constraints on the stellar pulsation model can be determined independently from the properties of the clusters 
\citep[see a recent excellent review by][]{pigulski2006}. 
One can get high-precision photometric results for open clusters because stars in open clusters are less crowded than those in globular clusters. 
Therefore open clusters are very important targets to study low amplitude pulsating stars \citep[e.g.][]{stello2006}.

Delta Scuti--type variable stars are A3$\sim$F0 type main-sequence or subgiant stars in the lower part of classical instability strip.
They have short pulsating periods from 0.02 to 0.3 days and amplitudes less than 1.0 mag \citep{breger1979}. 
Most of them have amplitudes less than 0.1 mag \citep{Rodriguez2001}. Many $\delta$ Scuti--type stars are discovered in intermediate-age 
open clusters because the main-sequence turn-off point of the clusters overlaps the $\delta$ Scuti instability strip. 
For example, \citet{arentoft2005} discovered 11 $\delta$ Scuti--type stars in the intermediate-age open cluster NGC 1817.

Eclipsing binary stars are found in open clusters as well. They are located above the single-star main-sequence in color-magnitude 
diagram of the clusters. Detached eclipsing binaries offer an opportunity to measure directly the stellar parameters such as mass, luminosity and radius.
Therefore, eclipsing binaries in open clusters can be used as distance indicator and to check stellar evolution theory \citep{paczynski1997}.
Age of the cluster versus relative incidence between long period detached eclipsing binaries and short period contact binaries
gives us very important constraint on dynamical evolution of binary systems \citep{rucinski1996}.
According to the most popular binary evolution model by \citet{huang1967} and \citet{vilhu1982}, W UMa--type binaries evolve into a contact configuration 
from initially detached systems by angular momentum loss via magnetic torque. Detached binaries with initial periods as long as 5$\sim$10 days 
may evolve into a contact configuration with orbital periods shorter than 1 day on a time-scale of a few Gyr. 
Observational results support the model, i.e., W UMa--type contact binaries are observed to be present in old open clusters
of about 4$\sim$5 Gyr \citep{kaluzny1993} but have not been observed to be present in open clusters younger than about 1 Gyr (\citealp[hereafter HSB]{hargis2005}; \citealp{rucinski1998}). 

As part of our long term project to survey variable stars in open clusters, extensive time-series CCD observations have been performed for 
intermediate-age open clusters using a 1.0m telescope at the Mount Lemmon Optical Astronomy Observatory (LOAO) in Arizona, USA. 
The primary goal of this project is to search for eclipsing binaries and short-period (less than a few days) pulsating variables,
and to study their physical properties in detail. From the extensive list of variable stars in the cluster, we can examine the possible 
relevance of characteristics of variable stars to cluster parameters. 
\citet{kang2007} detected a total of 41 variable stars in the intermediate-age open cluster NGC 2099
which was the first observing target of our project.

We selected M11 (NGC 6705; RA$_{\rm{J}2000.0}$= 18$^h$51$^m$04$^s$, DEC$_{\rm{J}2000.0}$= $-$06\arcdeg16\arcmin30\arcsec) as the second target. 
This cluster is one of the well established open clusters with intermediate-age of about 200$\sim$250 Myr 
\citep[][and references therein]{mermilliod1992,sung1999}. It is a rich and large cluster located
at a low Galactic latitude ($b=-2.8\arcdeg$). The most comprehensive study of variable stars 
in this cluster was recently performed by HSB. They surveyed the cluster center 
with a field of view of 13.7$\arcmin$$\times$13.7$\arcmin$ and detected 39 variables stars. 
Considering large radius of this cluster \citep[16\arcmin;][]{sung1999}, however, more extensive observations 
with wider area are needed to secure complete list of variable stars. 
In this paper, we present results of variable stars detected from our time-series observations 
with an wider field of view and relatively long time span. 

In Section 2, we present our observations and data analysis. Section 3 describes the physical properties of 
the pulsating variables and eclipsing binary stars detected from our observations. We discuss the cluster membership of 
the variable stars in Section 4. Summary and conclusion are given in Section 5.

\section{Observations and Data Reduction}
We carried out time-series observations of M11 on 18 nights in June 2004, using a 2K$\times$2K CCD camera attached to the 
LOAO 1.0m telescope in Arizona. The field of view of a CCD image is about 22.2$\times$22.2 arcmin$^2$, 
given a CCD plate scale of 0.64$\arcsec$ pixel$^{-1}$ at the f/7.5 Cassegrain focus of the telescope. 
A total of 1001 time-series images, i.e. 406 images with a long--exposure of 600 s and 595 images with a short--exposure of 60 s, 
were obtained with $V$-band filter to secure faint objects as well as bright ones. 
In order to minimize position dependent external errors \citep{frandsen1989}, 
we carefully controlled the telescope to keep the stars at fixed pixel positions on the CCD during our observing run. 
For the purpose of constructing $BV$ color-magnitude diagram of M11, additional $B$-band observations were made on 
one night of October 2004; two frames with exposure times of 1000 s and 100 s. During the whole observing run, 
the typical seeing disk of a star (FWHM) was about 2.2$\arcsec$. Figure 1 displays the observing CCD field of M11.

Instrumental signatures of each CCD frame were removed and calibrated using the bias, dark, and flat field frames, 
with the aid of the IRAF package CCDRED. We obtained instrumental magnitudes of stars from the empirical 
point--spread function (PSF) fitting method in the IRAF package DAOPHOT \citep{stetson1987,massey1992}.

We applied an ensemble normalization technique \citep{gilliland1988} in order to normalize instrumental magnitudes 
of CCD frames, following the same procedure used by \citet{kim2001}. Using this technique we corrected for the color and position 
dependent effects of the observation system and atmospheric differential extinction for all CCD frames. 
A few tens of bright stars with wide color range, selected from \citet{sung1999}, were used as the secondary standard stars.

\section{Variable Stars}
We carefully examined the light variations of about 32,000 stars by visual inspection. 
Saturated stars and the stars located at the edge of CCD frames were excluded. 
A total of 82 variable stars were discovered: 13 pulsating stars and 69 eclipsing binaries. 
A finding chart of the variable stars is shown in Figure 1 and their observational properties are presented in Table 1 and 2.
We derived the periods of pulsating variable stars using the discrete Fourier analysis \citep{scargle1989,kim2001}.
Table 3 summarizes results of the multiple frequency analysis. 
The phase matching method \citep{hoffmeister1985} was applied for eclipsing binaries to estimate their orbital periods. 

\subsection{Previously Known Variable Stars}
Six variable stars had been known in the open cluster M11 before HSB's study.
BS Scuti was located outside our observing field and a bright variable V369 Scuti ($V$ = 9.35 mag) was saturated in our data.
IT Scuti has been known to be a slow irregular variable but we could not detect its light variations.
We could not find out any light variations for the other three suspected variables, NSV11410, NSV11402 and NSV24615.

HSB detected a total of 39 variable stars including six $\delta$ Scuti--type stars (HV1$\sim$HV6), 
17 W UMa--type variables (HV8, HV10$\sim$HV25), 14 detached eclipsing binaries (HV26$\sim$HV39), one irregular variable (HV7), 
and one unclassified candidate variable (HV9).
We could identify all of these variable stars in our observations.

Power spectra of six $\delta$ Scuti--type pulsating stars are shown in Figure 2.
Two stars of HV2 and HV6 show a dominant frequency with relatively large power; $f_1$=18.271 c/d (cycles per day) for HV2 
and $f_1$=12.373 c/d for HV6. In particular, the large power of HV6 is characteristic of high amplitude $\delta$ Scuti--type variable stars,
as noted by HSB, which differ from most $\delta$ Scuti--type pulsators with small amplitudes 
\citep[see][and references therein]{rodriguez1996,rodriguez2004}.
Pulsating periods of the four variables HV1, HV2, HV3, and HV6 are coincident with those of HSB.
But our periods for HV4 and HV5 are different from HSB's results, probably due to
the low amplitudes and multiple periodicities; our data show a weak signal near 24 c/d for HV5 which corresponds to the period of 0.04154 days by HSB.
Low frequencies less than 5 c/d with low amplitudes about 5$\sim$8 mmag ($f_2$ for HV4 and $f_1$ for HV5) may be originated from
the slow variations of our observation system and/or atmospheric conditions.

HSB has classified HV7 (their ID 220) as an irregular variable to have both long and short timescale
light variations. As shown in Figure 3, we confirmed the long timescale variations of about 0.633 days but failed to
detect the very short timescale variations of about 10 minutes.
Comparing with the other variables with similar brightness such as HV8, phase diagram of HV7 shows
rather large scatter over the whole phase, implying existence of multiple frequencies.
HSB also noted that HV7 showed different light variations from night to night.
On the basis of the variable period, multiple periodicities, light curve shape, and the position on a color-magnitude diagram located within
the $\delta$ Scuti instability strip (see Figure 9), we suggest that HV7 is a $\gamma$ Doradus--type pulsating star.
In order to make a confirmation of the $\gamma$ Doradus--type pulsation and rule out the possibility of the binary or rotation effect, high-resolution time-series spectroscopic observations of HV7 would be required.

We confirmed all the 17 W UMa--type binary stars detected by HSB.
Orbital periods obtained in our study are in good agreement with HSB's results, except for HV21.
Period difference of HV21 may be resulted from the well-known 1.0 c/d alias effect for the data obtained at single observatory.
In addition to these binaries, our data showed very clearly that HV9 (their ID 708) is a W UMa--type binary.
HSB could not classify the variable type of HV9 because it was near the edge of CCD frame and 
then they had obtained data on only two nights.
Figure 4 displays phase diagrams of these 18 W UMa--type binary stars. 

HSB identified 14 detached eclipsing binary stars. Among these binaries, they detected multiple eclipses
for six systems (HV26, HV27, HV28, HV29, HV31, and HV32) and then could determine their orbital periods. 
From our more extensive data-set with longer time span, we obtained full light curves (left panels of Figure 6) for these six
binaries and determined more accurate periods and epochs (listed in Table 1). 
In the case of HV29, HSB determined the orbital period of 5.62050 days.
However, based on our complete light curve, we obtained a half orbital period (2.8060 days) for this system.

HSB detected only one or two eclipses for remaining eight systems (HV30, HV33, HV34, HV35, HV36, HV37, HV38, and HV39) 
and so could not determine their orbital periods. 
Among these eight systems, we could determine the period of 4.460 days and epoch for HV30. 
For four systems (HV33, HV34, HV35, and HV38), only one or two eclipsing features were detected in our data.
We tried to determine their orbital periods combining with HSB's epochs and ours, but failed due to large separation of epochs.
For two systems of HV36 and HV39, our data did not show eclipsing feature.
HV37 (their ID 4804) showed slow variations, the same as HSB, 
but we could not determine its variable type. Figure 7 displays light variations of these seven eclipsing binaries which we could not estimate orbital periods.

\subsection{New Pulsating Variables}
We discovered six new pulsating variable stars in our observing field. 
On the basis of pulsating periods and positions on the color-magnitude diagram of the cluster, we classified five 
$\delta$ Scuti--type stars (KV1$\sim$KV5) and one $\gamma$ Doradus--type star (KV6).
The signal to noise amplitude ratio (S/N) greater than 4.0 \citep{breger1993} was used as a detection criterion for pulsating frequency.
The results of frequency analysis for these pulsating stars are listed in Table 3.

As shown in the right panels of Figure 2, KV2 and KV5 are high amplitude pulsating variables with large powers.
We checked the period ratio for these high amplitude $\delta$ Scuti--type stars and found that the ratio of $f_1$/$f_2$ = 0.781 $\pm$ 0.001 for KV2 is
similar to that of the theoretical radial modes P$_1$/P$_0$ = 0.761 \citep[for a model with Y=0.28, Z=0.02, M=1.7M$_\odot$, T$_{eff}$=7,000K, and L=15L$_\odot$;][]{breger1979}.
It indicates that KV2 may be excited in two radial modes, i.e. fundamental (P$_0$ for $f_1$) and first-overtone radial modes (P$_1$ for $f_2$).
KV5 seems to be a mono-periodic high amplitude $\delta$ Scuti--type star.

The other three stars (KV1, KV3, and KV4) show a few weak powers. They are inside or near the $\delta$ Scuti--type instability strip. 
We detected five frequencies for KV1. Its period ratios of $f_1$/$f_2$ = 0.617 $\pm$ 0.001 and $f_1$/$f_4$ = 0.520 $\pm$ 0.001 are nearly the same as those of 
theoretical radial modes P$_2$/P$_0$ = 0.616 and P$_3$/P$_0$ = 0.521 \citep{breger1979}, respectively, indicating that KV1 is excited in 
fundamental (P$_0$ for $f_1$), second-overtone (P$_2$ for $f_2$) and third-overtone (P$_3$ for $f_4$) radial modes.
The frequency $f_3$ = 3.003 c/d seems to be the 1.0 c/d alias effect of a combination frequency, $f_2$ $-$ $f_1$ $-$ 1.0 = 2.984 c/d, and
the other frequency $f_5$ may be a non-radial mode.
We detected two frequencies for KV3 and only one frequency for KV4.

KV6 has a long period of 0.9079 days which is comparable with that of $\gamma$ Doradus--type pulsating stars \citep{kaye1999}.
On the color-magnitude diagram (Figure 9), it is located outside the red edge of $\delta$ Scuti-instability strip and 
near the red edge of $\gamma$ Doradus-instability strip \citep{handler2002}.
Therefore, we suggest that KV6 is a $\gamma$ Doradus--type pulsating star.
The phase diagram of KV6 displays in the right panel of Figure 3.

\subsection{New Eclipsing Binaries}
We discovered new 22 W UMa--type binary stars (KV7$\sim$KV28). We obtained complete phase coverage for these binary systems,
except for KV21 which shows incomplete secondary minimum due to its orbital period of about 0.5 days. 
Figure 5 shows various shapes of the light curves of these systems with different amplitudes. 
Two systems of KV15 and KV16 show slightly different brightness between two maxima,
which gives a hint of the existence of spots on the stellar surface \citep{wilson1994}.
In several systems (e.g. KV10), the depths of two minima are appeared to be different, meaning significant temperature differences
between primary and secondary stars.

We discovered six new detached eclipsing binaries (KV29$\sim$KV34) in our observing field. 
Phase diagrams of these systems are presented in the right panels of Figure 6 (see also Table 2). Most of these systems have flat maxima to show 
a shape typical of detached binaries. Two systems of KV32 and KV33 show slightly round maxima in their phase diagrams 
indicating that they have relatively small separation between component stars and one of the component fills its Roche lobe. 
This is supported by their short orbital periods of about 0.65 days.

We also detected nine new eclipsing binary systems, KV35$\sim$KV43 (see Figure 8 and Table 2), but could not determine their ephemeris due to our limited data-set. They seem to have long orbital periods.
From the repeated occurrence of the minima in their light curves, we estimate that the orbital periods are about 6 days for KV36 
and about 2 days for KV35 and KV43.

\section{Color$-$Magnitude diagram and Membership}
$BV$ color-magnitude diagram of M11 is shown in Figure 9. The thick solid line represents the empirical zero-age main-sequence (ZAMS) 
from \citet{sungbessel1999}. The thin solid line is the theoretical isochrone of \citet{girardi2000} with a solar metal abundance (Z=0.019)
and an age of log $t_{\rm{age}}=8.35$, adopting the $E(\bv)$ = 0.428 and $(V-M_{V})_0$ = 11.55 from \citet{sung1999}. 
Long dashed line is the possible equal-mass binary sequence to the isochrone as a guidance of the membership of detected eclipsing binaries. 
Solid bars, nearly perpendicular to the ZAMS, represent the $\delta$ Scuti instability strip \citep{breger1979}.

In Figure 9, we show the positions of 72 variable stars with different symbols for different type of variables; 
star symbols for pulsating stars, filled circles for W UMa--type binaries and open triangles for detached systems. 
In the figure, 9 faint variables and one star (KV32) near the edge of the CCD chip are excluded, because we could not obtain their $B$ magnitudes.
Since the color-magnitude diagram of M11 has very large contamination by field stars \citep{mathieu1984, brocato1993, sung1999}, it is not easy
to distinguish cluster member stars from the field population. The sequence of the ``blue'' field star population ($\bv<1.2$ mag) is overlapped
with the cluster main-sequence at about $V>15$ mag \citep[see Figure 3 and Figure 6 of][]{sung1999}, which severely hampers the identification 
of member variable stars detected in our study. The ``red'' field population is also a main contaminator of membership estimation
for the faint and red variable stars with $V>17$ mag and $\bv>1.2$ mag. \citet{mcnamara1977} and \citet{su1998} provided probabilities of membership 
for bright stars in the field of M11 based on the relative proper motions.    

All of the 13 pulsating stars, except for two fainter ones KV5 and HV6, are located inside or near the $\delta$ Scuti instability strip, suggesting that they may be probable member stars in the cluster. 
Available membership probability information of five $\delta$ Scuti--type stars (KV1: 99\%, HV2: 98\%, HV3: 96\%, HV4: 83\%, and HV5: 80\%) 
from \citet{mcnamara1977} supports the possible membership of these stars. 
On the other hand, \citet{mcnamara1977} found a zero membership probability for two $\delta$ Scuti--type stars (HV1 and KV2) 
and a $\gamma$ Doradus--type star (HV7).
In the case of two faint variables KV5 and HV6, although membership probabilities are not available from \citet{mcnamara1977},
their faintness implies that they are field $\delta$ Scuti--type variable stars.

According to the membership information of \citet{mcnamara1977}, three bright ($V<$ 14.0 mag) binary stars (KV29, KV35, and KV37) 
show high ($P_\mu$ $>$ 97\%) membership probabilities. However, some other bright binary stars have low membership probabilities; 0\% for HV8 and 5\% for KV36.
Although many fainter ($V>$ 15.0 mag) binary stars are located in the region between empirical ZAMS and equal-mass binary sequence, 
it is difficult to say with certainty whether these stars are cluster members due to the large contamination of the field star population.
Binary stars fainter than the main-sequence or brighter than the equal-mass binary sequence are likely non-members of the cluster. Futhermore, as displayed in Figure 1, most of contact binaries are located outside of the cluster half-mass radius \citep[$r$=4.5$\arcmin$;][]{mathieu1984} and then seem to be non-members \citep{hargis2005}.

\section{Summary \& Conclusion}
From our extensive photometric variability survey of the intermediate-age open cluster M11,
the number of detected variable stars have doubled from previous studies, being total 82 stars.
It consists of 13 pulsating stars, 40 W UMa--type eclipsing binaries, 13 detached eclipsing binaries, and 16 eclipsing binaries with long period.

Eleven out of 13 pulsating variables are within or near the $\delta$ Scuti instability strip so that they seem to be probable members of the cluster;
five stars (HV2, HV3, HV4, HV5, and KV1) in the strip have high membership probability of $P_\mu \ge$ 80\% but three stars (HV1, HV7, and KV2) are
non-members with $P_\mu =$ 0\%.
Among a few hundreds of stars in the strip, only seven stars (excluding two non-members and two $\gamma$ Doradus--type stars) 
turned out to be $\delta$ Scuti--type pulsating variables.
The incidence of less than 5\% is much lower than the cases of field stars \citep[being about 30\%;][]{breger1979, wolff1983} and of
some open clusters \citep{frandsen1996,horan1979}. On the other hand, the low incidence of $\delta$ Scuti--type stars 
in M11 is similar to the cases of many other open clusters \citep{kang2007,viskum1997}.

We detected 13 detached and 40 contact eclipsing binaries in the field of M11.
Several detached binaries are located between the ZAMS and the equal-mass binary sequence on the color-magnitude diagram.
Four of them (HV33, KV29, KV35, and KV37) have high membership probability of $P_\mu \ge$ 80\%.
We found two W UMa--type contact binaries (HV10 and HV16) to be possible cluster members, i.e., they are inside the cluster half-mass radius and are located between the ZAMS and the binary sequence. However, since the membership information of HV10 and HV16 is not available, we could not ensure whether the open cluster M11 with intermediate-age ($\sim$200 Myr) can contain contact binaries.
It is of importance to find out the existence of contact binaries in intermediate-age open clusters regarding the understanding of the dynamical evolution of binary systems. Therefore, further proper motion and/or radial velocity survey for contact binaries in the field of M11 
should be required to secure their cluster memberships.

In conclusion, because many pulsating stars and eclipsing binaries in the region of M11 are probable members of the cluster, 
we believe that the cluster M11 is one of the best observing targets for asteroseismology studies as well as for investigating dynamical evolution of binary systems.

We thank J.-W. Lee and Santabrata Das for their careful reading and comments. We greatly appreciate an anonymous referee for the valuable suggestions. This research has made use of the WEBDA database, operated at the Institute for Astronomy of the University of Vienna. This study was financially supported by research fund of Chungnam National University in 2005.

\clearpage
\begin{figure}
 \includegraphics[]{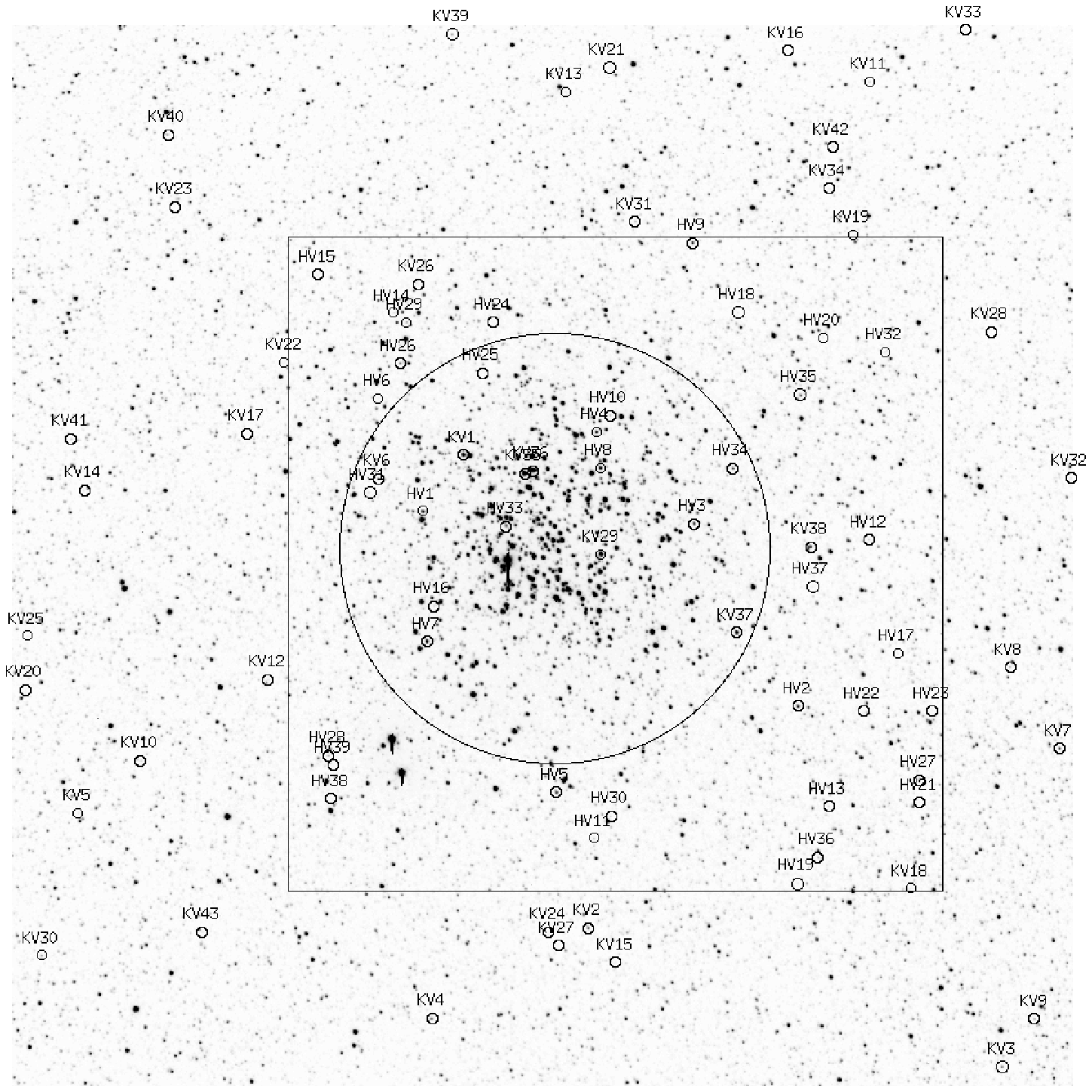}
 \caption{Observed CCD field (22.2$\times$22.2 arcmin$^2$) of M11. Identification of variable stars are also marked. Large circle and large box represent the cluster half-mass radius \citep[$r$=4.5$\arcmin$;][]{mathieu1984} and observation field of \citet{hargis2005}, respectively. North is up and east is to the left.
 \label{Fig1}}
\end{figure}

\clearpage
\begin{figure}
  \includegraphics[]{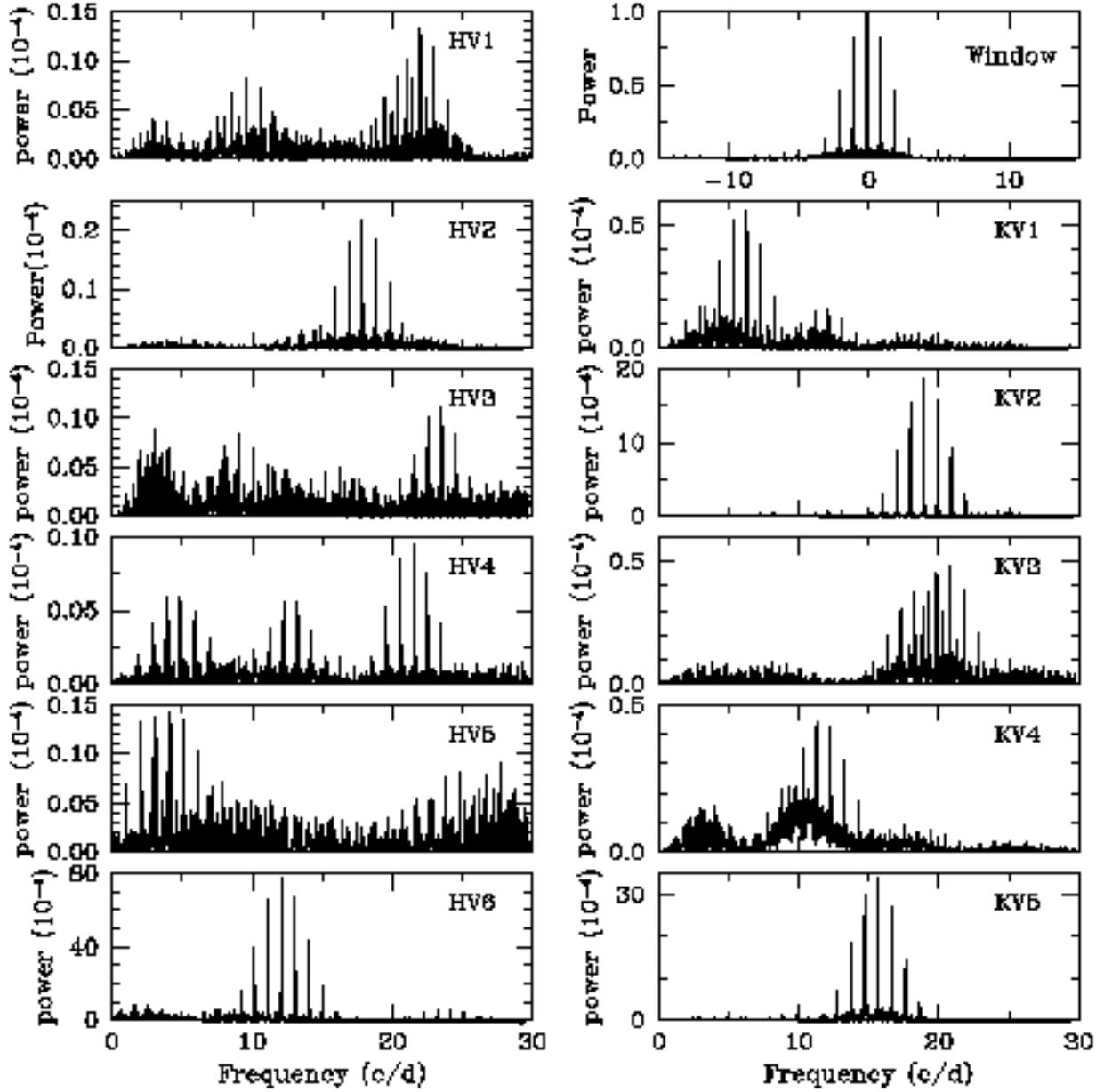}
 \caption{(Left) Power spectra of six previously known $\delta$ Scuti--type stars and (right) five newly detected ones. 
          Top right panel shows a spectral window spectrum.
\label{Fig2}}
\end{figure}

\clearpage
\begin{figure}
  \includegraphics[]{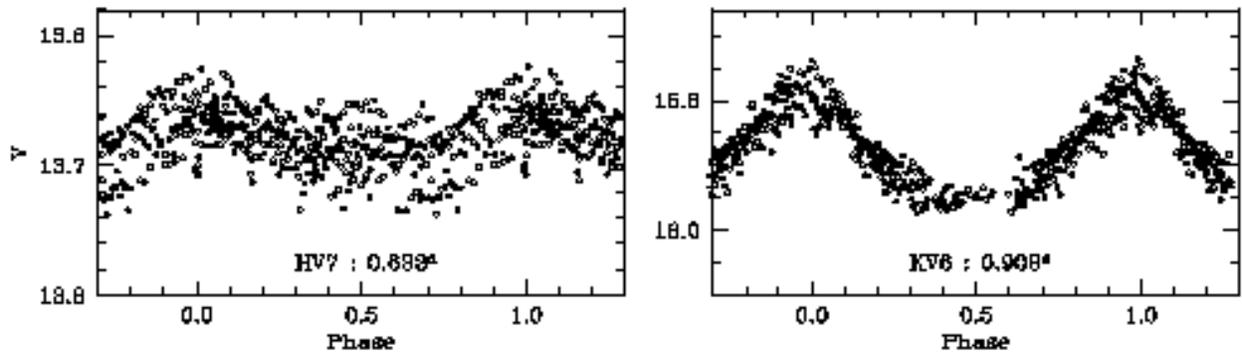}
 \caption{Phase diagrams of two $\gamma$ Doradus--type stars.
 \label{Fig3}}
\end{figure}

\clearpage
\begin{figure}
\includegraphics[]{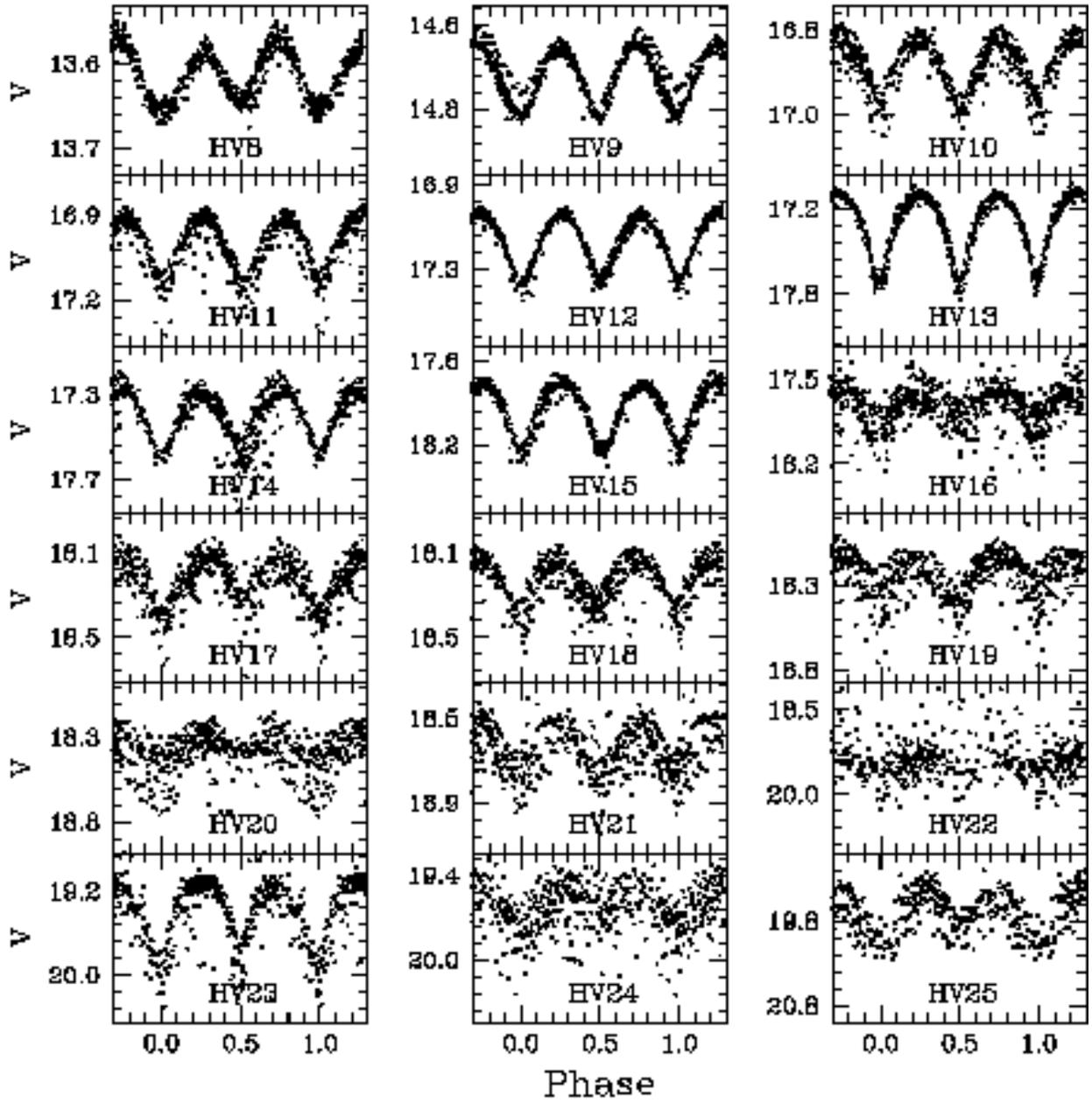}
\caption{Phase diagrams for 18 previously known W UMa--type variable stars.
\label{Fig4}}
\end{figure}

\clearpage
\begin{figure}
\includegraphics[]{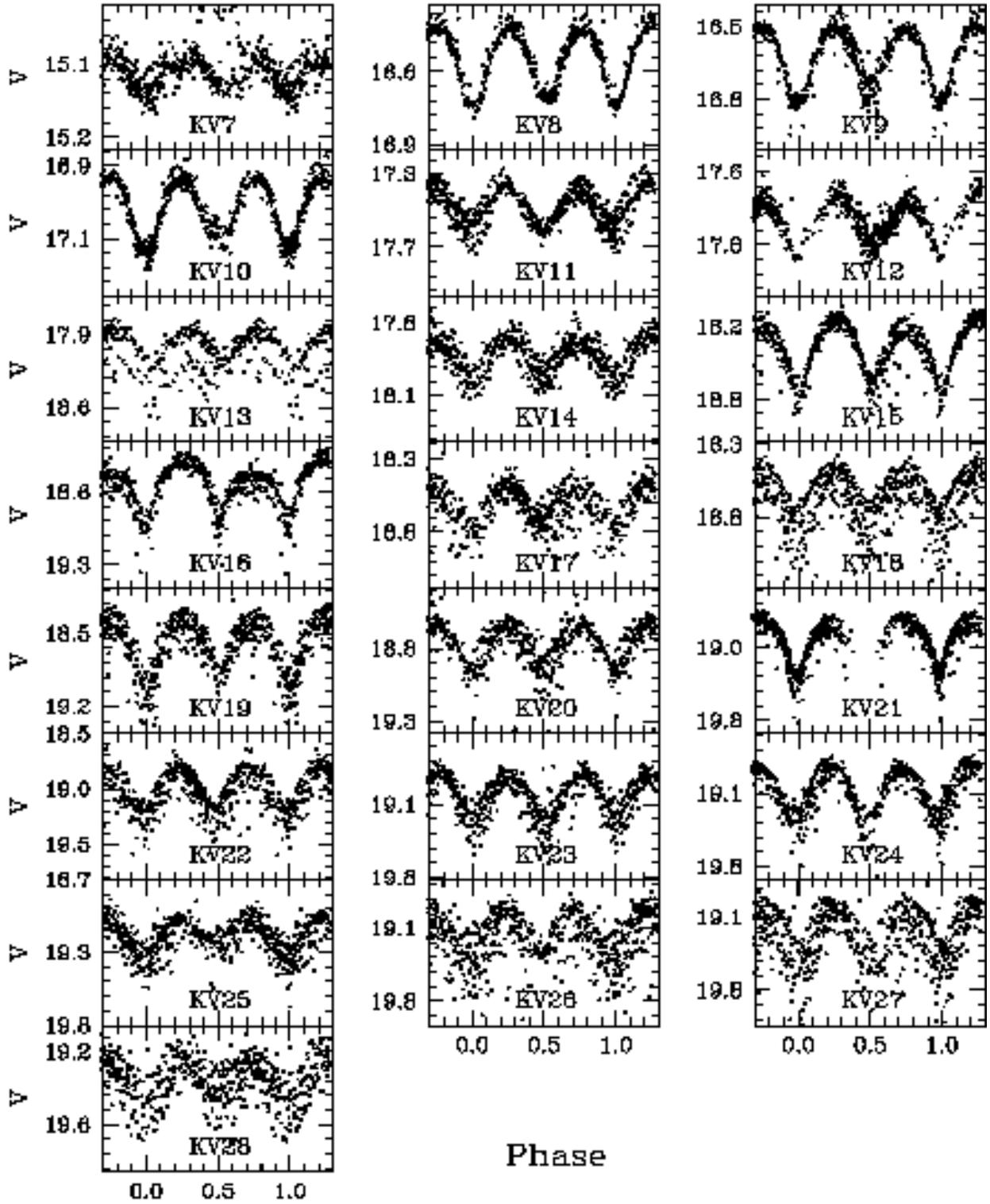}
\caption{Phase diagrams for 22 new W UMa--type variable stars detected in our study.
\label{Fig5}}
\end{figure}

\clearpage
\begin{figure}
  \includegraphics[]{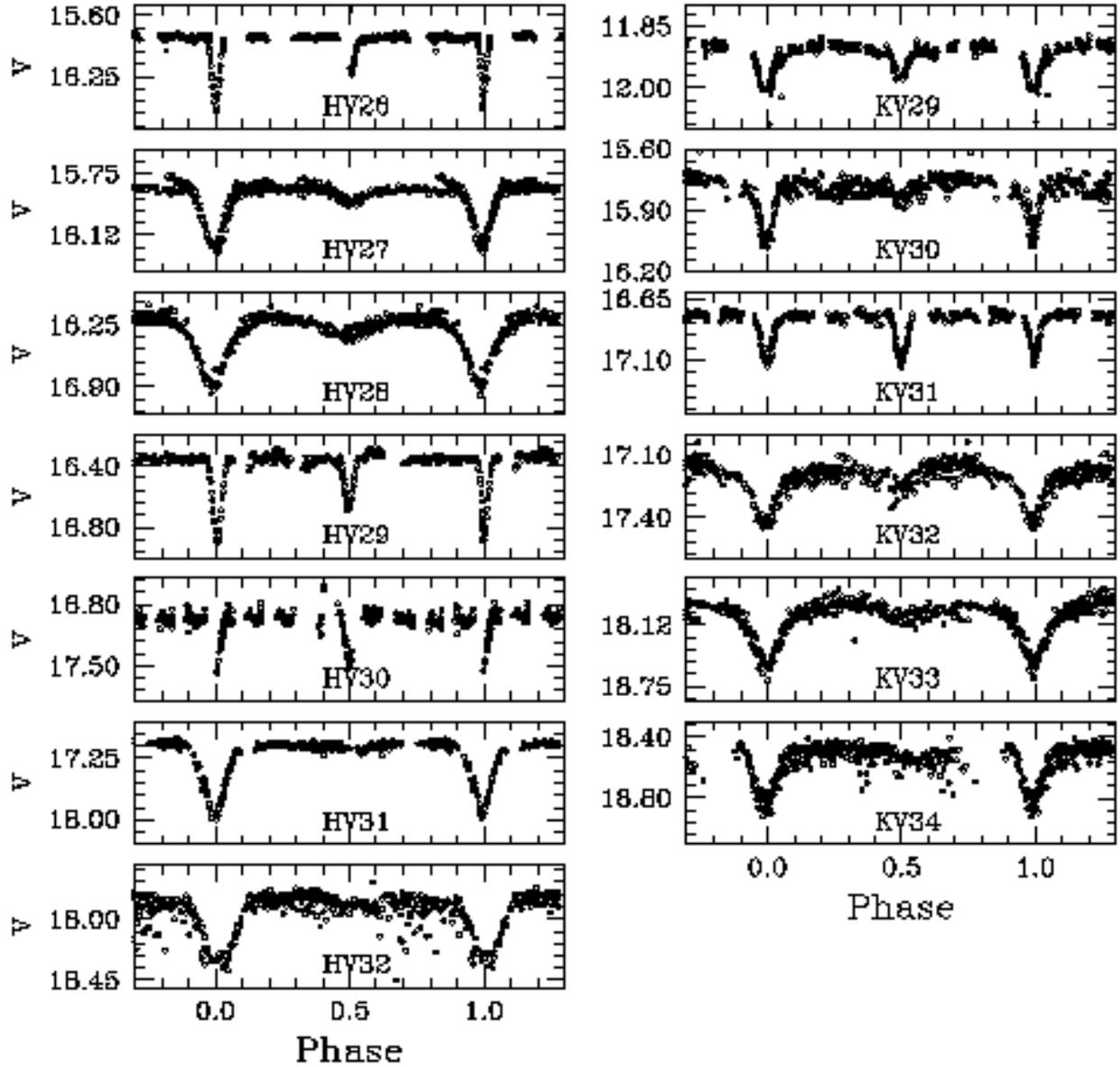}
 \caption{Phase diagrams for 13 detached binary systems; (left) seven previously known systems and (right) six new systems detected in our study.
\label{Fig6}}
\end{figure}

\clearpage
\begin{figure}
\includegraphics[]{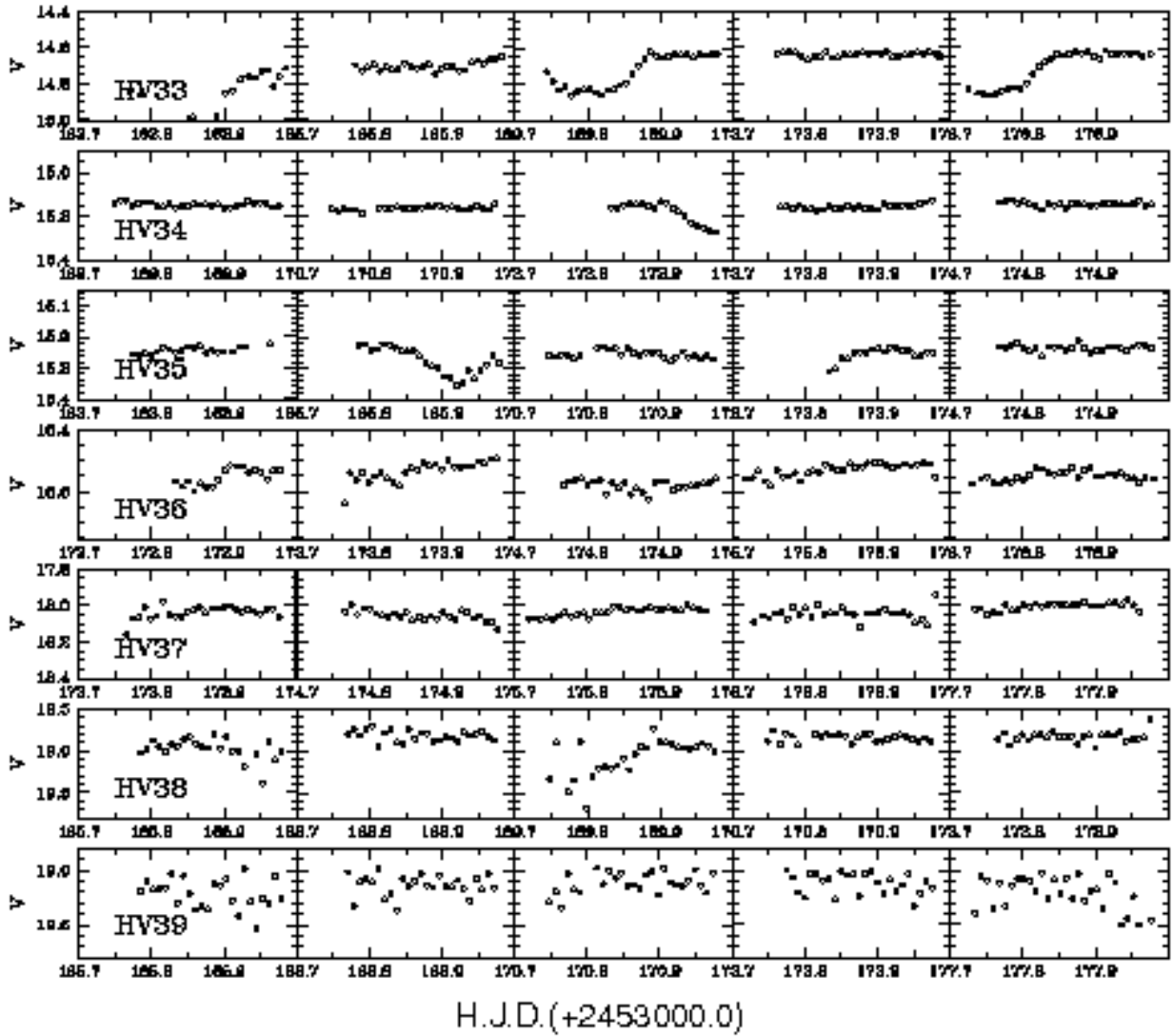}
\caption{Light curves of 7 previously reported long period eclipsing binary systems in \citet{hargis2005}.
\label{Fig7}}
\end{figure}

\clearpage
\begin{figure}
\includegraphics[]{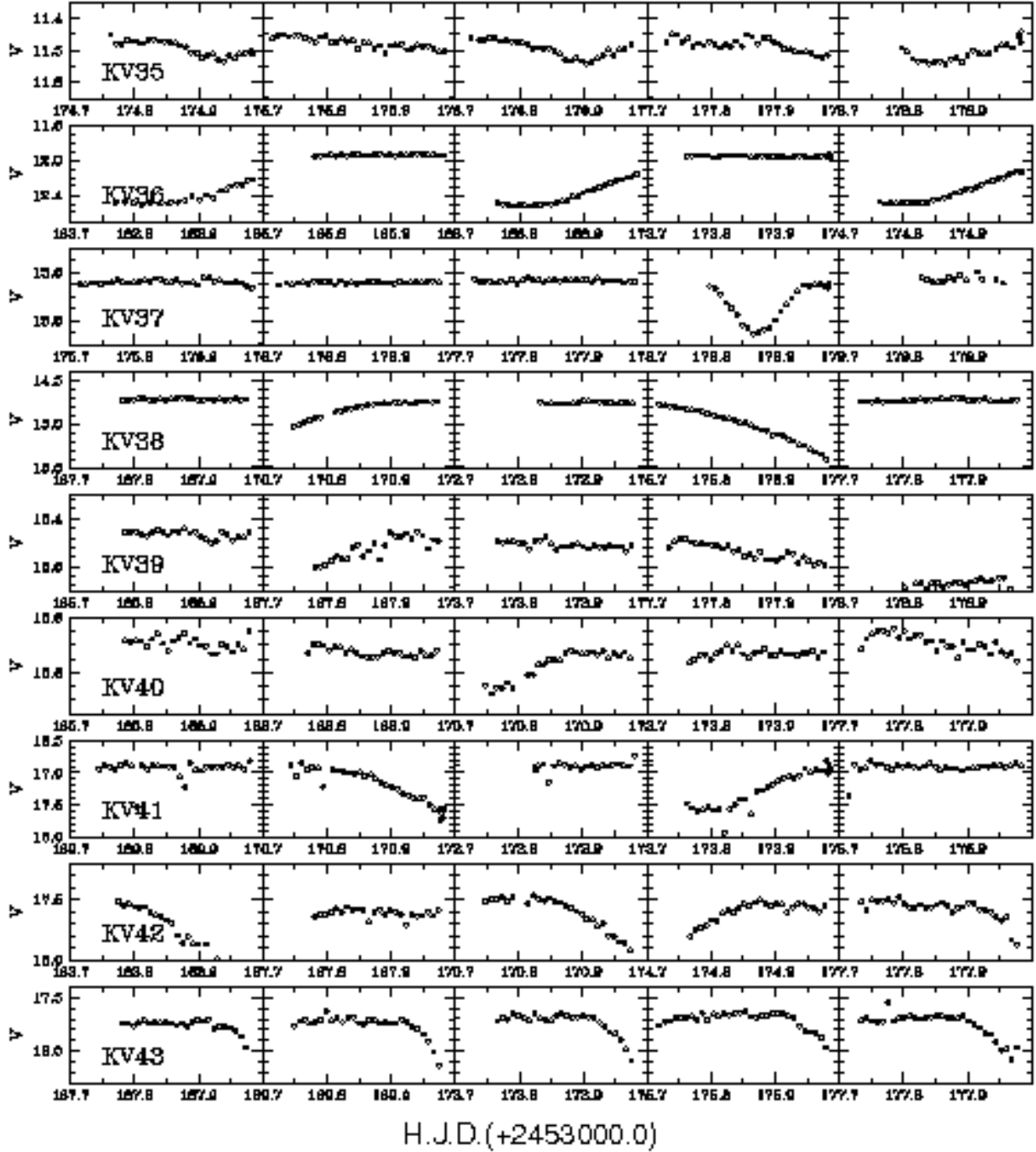}
\caption{Light curves of 9 new long period eclipsing binary systems detected in our study.
\label{Fig8}}
\end{figure}

\clearpage
\begin{figure}
\includegraphics[]{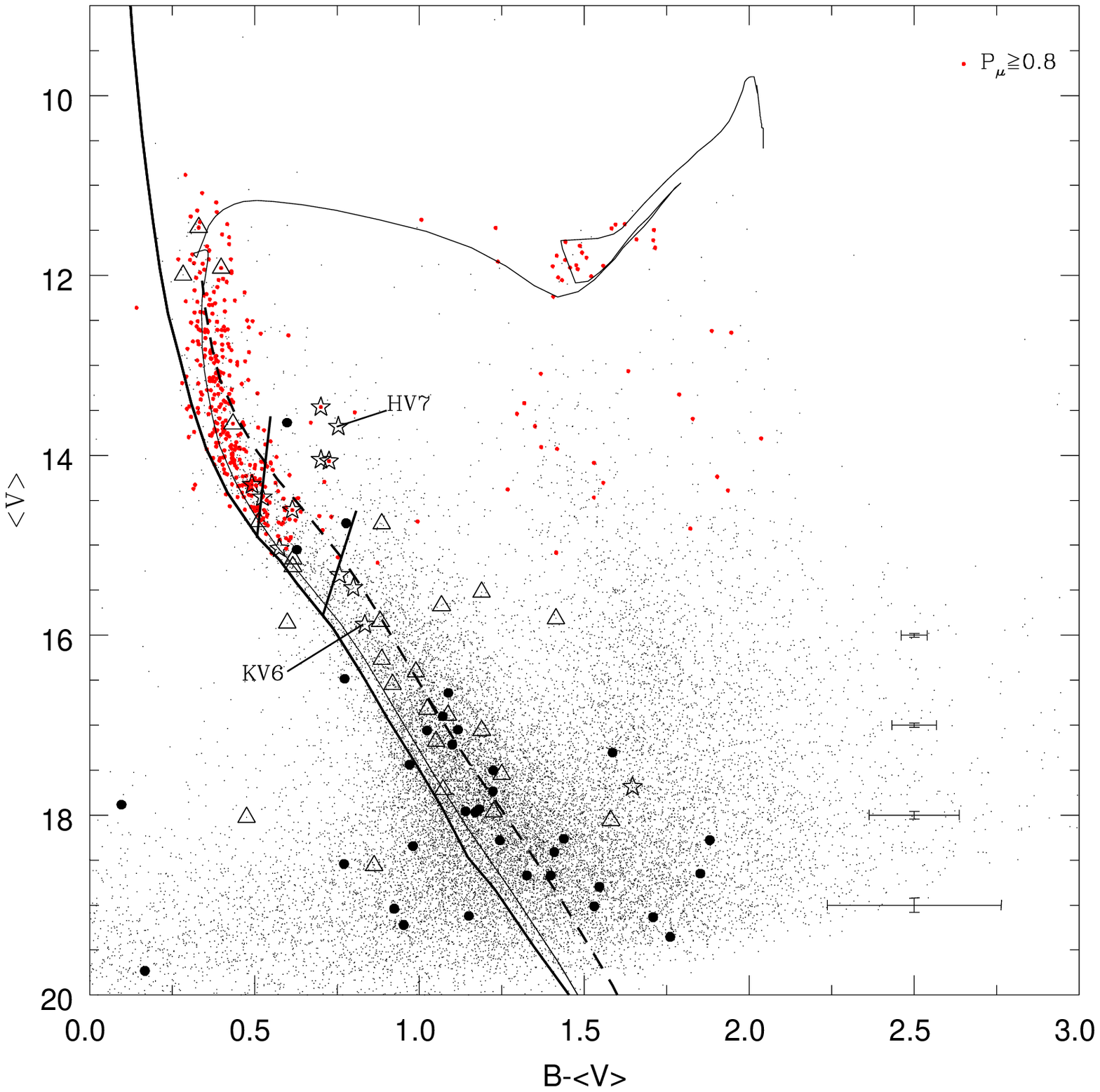}
\caption{Color-magnitude diagram of M11. The thick solid line is the empirical ZAMS from \citet{sungbessel1999}. 
The thin solid line represents the theoretical isochrone of \citet{girardi2000} with Z = 0.019 and log $t_{\rm{age}}$ (year)=8.35.
The long dashed line is the theoretical equal-mass binary sequence to the isochrone. Solid bars represent the $\delta$ Scuti instability strip \citep{breger1979}. Large dots represent probable 
member stars with $P_\mu$ (membership probability) $\geq$ 0.8 from \citet{mcnamara1977}. Star symbols, filled circles, and  open triangles are pulsating stars, W UMa--type binaries, and detached systems, respectively.
\label{Fig9}}
\end{figure}

\clearpage
\begin{table}
\footnotesize
\caption{Basic Information for 39 Variables previously detected by \citet{hargis2005}. \label{Tab1}}
\begin{tabular}{ccccccccccc}
\hline
  ID &  ID$_{Hargis}$  &ID$_{WEBDA}$& P & R.A.(J2000.0)  & Decl.(J2000.0)  & $V$ & \bv& Period &  Epoch (H.J.D)  & Type \\
    &    & & \% & $^h$ $^m$ $^s$   & \arcdeg\ \arcmin\ \arcsec &  mag   &    & (days)   &  (+2453000.0) &\\\hline

HV1   & 331  &  676 &  0  &18 :51 :15.1 & $-$06 :15 :43.8 & 14.048 & 0.702 & 0.04504   &165.950 &$\delta$ Scuti\\
HV2   & 320  & 1711 & 98  &18 :50 :43.6 & $-$06 :19 :47.5 & 14.064 & 0.726 & 0.05473   &161.828 &$\delta$ Scuti\\
HV3   & 536  & 1531 & 96  &18 :50 :52.3 & $-$06 :15 :59.0 & 14.327 & 0.492 & 0.04209   &178.563 &$\delta$ Scuti\\
HV4   & 614  & 1237 & 83  &18 :51 :00.5 & $-$06 :14 :04.3 & 14.466 & 0.524 & 0.04603   &165.983 &$\delta$ Scuti\\
HV5   & 619  & 1097 & 80  &18 :51 :03.9 & $-$06 :21 :35.7 & 14.607 & 0.615 & 0.03576   &165.940 &$\delta$ Scuti\\
HV6   & 6870 &  $-$ &$-$  &18 :51 :18.8 & $-$06 :13 :22.2 & 19.001 &  $-$  & 0.08082   &178.563 &$\delta$ Scuti\\
HV7   & 0220 &  683 &  0 &18 :51 :14.7 & $-$06 :18 :26.5 & 13.680 & 0.754 & 0.63315  & 161.850 &$\gamma$ Doradus\\
HV8   & 0243 & 1258 &  0 &18 :51 :00.2 & $-$06 :14 :49.4 & 13.636 & 0.599 & 0.869565 & 163.295 &W UMa\\
HV9   & 0708 & 1521 &  0 &18 :50 :52.4 & $-$06 :10 :07.2 & 14.754 & 0.778 & 0.676796 & 161.815 &W UMa\\
HV10  & 2740 &  $-$ &$-$&18 :50 :59.4 & $-$06 :13 :43.3 & 16.900 & 1.070 & 0.394555 & 162.790 &W UMa\\
HV11  & 3064 &  $-$ &$-$&18 :51 :00.7 & $-$06 :22 :33.3 & 17.060 & 1.023 & 0.441209 & 161.732 &W UMa\\
HV12  & 3250 &  $-$ &$-$&18 :50 :37.6 & $-$06 :16 :19.0 & 17.216 & 1.100 & 0.352112 & 161.785 &W UMa\\
HV13  & 3025 &  $-$ &$-$&18 :50 :41.0 & $-$06 :21 :53.1 & 17.305 & 1.585 & 0.441989 & 161.853 &W UMa\\
HV14  & 3597 &  $-$ &$-$&18 :51 :17.6 & $-$06 :11 :35.6 & 17.440 & 0.970 & 0.473709 & 162.880 &W UMa\\
HV15  & 5274 &  $-$ &$-$&18 :51 :23.9 & $-$06 :10 :47.9 & 17.933 & 1.181 & 0.34662  & 161.795 &W UMa\\
HV16  & 3408 &  $-$ &$-$&18 :51 :14.2 & $-$06 :17 :42.1 & 17.959 & 1.140 & 0.25532  & 166.058 &W UMa\\
HV17  & 4724 &  $-$ &$-$&18 :50 :35.2 & $-$06 :18 :41.1 & 18.263 & 1.438 & 0.417188 & 162.862 &W UMa\\
HV18  & 5710 &  $-$ &$-$&18 :50 :48.6 & $-$06 :11 :34.5 & 18.279 & 1.880 & 0.462877 & 162.676 &W UMa\\
HV19  & 4779 &  $-$ &$-$&18 :50 :43.6 & $-$06 :23 :29.1 & 18.281 & 1.244 & 0.38382  & 162.868 &W UMa\\
HV20  & 5480 &  $-$ &$-$&18 :50 :41.5 & $-$06 :12 :06.1 & 18.411 & 1.409 & 0.43009  & 165.910 &W UMa\\
HV21  & 6647 &  $-$ &$-$&18 :50 :33.4 & $-$06 :21 :47.7 & 18.670 & 1.398 & 0.547600 & 165.000 &W UMa\\
HV22  & 6805 &  $-$ &$-$&18 :50 :38.1 & $-$06 :19 :53.9 & 19.221 & 0.952 & 0.34516  & 161.950 &W UMa\\
HV23  & 8066 &  $-$ &$-$&18 :50 :32.3 & $-$06 :19 :53.3 & 19.408 & $-$ & 0.42180  & 163.585 &W UMa\\
HV24  & 8641 &  $-$ &$-$&18 :51 :09.2 & $-$06 :11 :46.7 & 19.614 & $-$ & 0.45950  & 165.560 &W UMa\\
HV25  & 8146 &  $-$ &$-$&18 :51 :10.1 & $-$06 :12 :50.2 & 19.732 & 0.168 & 0.29357  & 165.786 &W UMa\\
HV26  & 1340 & 7432 &$-$&18 :51 :17.0 & $-$06 :12 :38.4 & 15.847 & 0.882 & 3.79130  & 163.865 &EA\\
HV27  & 1583 & 1917 & 38 &18 :50 :33.4 & $-$06 :21 :19.2 & 15.864 & 0.600 & 1.11775  & 177.720 &EA\\
HV28  & 1814 &  482 & 11 &18 :51 :23.0 & $-$06 :20 :49.9 & 16.268 & 0.886 & 0.70519  & 178.840 &EA\\
HV29  & 1938 & 7410 &$-$&18 :51 :16.5 & $-$06 :11 :47.1 & 16.408 & 0.990 & 2.80600  & 177.200 &EA\\
HV30  & 2406 &  $-$ &$-$&18 :50 :59.3 & $-$06 :22 :05.3 & 17.060 & 1.189 & 4.460    & 165.760 &EA\\
HV31  & 3096 &  $-$ &$-$&18 :51 :19.6 & $-$06 :15 :20.4 & 17.181 & 1.050 & 1.65174  & 166.560 &EA\\
HV32  & 4678 &  $-$ &$-$&18 :50 :36.3 & $-$06 :12 :23.9 & 17.962 & 1.225 & 0.72511  & 177.980 &EA\\
HV33  & 0729 &  906 & 86 &18 :51 :08.1 & $-$06 :16 :01.8 & 14.738 & 0.514 & $-$    & $-$ &EA\\
HV34  & 0977 & 1596 & 52 &18 :50 :49.1 & $-$06 :14 :49.8 & 15.151 & 0.617 & $-$    & $-$ &EA\\
HV35  & 1026 & 1715 &  7 &18 :50 :43.4 & $-$06 :13: 17.7 & 15.237 & 0.617 & $-$    & $-$ &EA\\
HV36  & 2119 & 5594 &$-$&18 :50 :41.9 & $-$06 :22 :56.8 & 16.547 & 0.919 & $-$    & $-$ &EA\\
HV37  & 4804 &  $-$ &$-$&18 :50 :42.4 & $-$06 :17 :17.7 & 18.062 & 1.581 & $-$    & $-$ &Slow(?)\\
HV38  & 7467 &  $-$ &$-$&18 :51 :22.8 & $-$06 :21 :43.4 & 18.874 & $-$ & $-$    & $-$ &EA\\
HV39  & 7522 &  $-$ &$-$&18 :51 :22.6 & $-$06 :21 :02.4 & 19.163 & $-$ & $-$    & $-$ &EA\\\hline
\end{tabular}
\tablecomments{For $\delta$ Scuti--type pulsating stars, the period and epoch are based on a dominated frequency for each variable. The fourth column represents the membership probability by \citet{mcnamara1977}.}
\end{table}

\clearpage
\begin{table}
\footnotesize
\caption{Basic Information for 43 Variable Stars newly detected in Our Study. \label{Tab2}}
\begin{tabular}{cccccccccc}
\hline
  ID &ID$_{WEBDA}$& P & R.A.(J2000.0)  & Decl.(J2000.0) & $V$ & \bv& Period &  Epoch (H.J.D)  & Type\\
        &     & \% & $^h$ $^m$ $^s$   & \arcdeg\ \arcmin\ \arcsec &  mag&  & (days)& (+2453000.0)&\\\hline

 KV1   &  764 & 99  &18 :51 :11.7 & $-$06 :14 :33.1 & 13.464 & 0.701 & 0.15544   &166.010  & $\delta$ Scuti\\
 KV2   & 1217 &  0  &18 :51 :01.2 & $-$06 :24 :25.2 & 15.035 & 0.576 & 0.05158   &161.762  & $\delta$ Scuti\\
 KV3   & 4928 &$-$  &18 :50 :26.4 & $-$06 :27 :19.3 & 15.330 & 0.757 & 0.04759   &165.960  & $\delta$ Scuti\\
 KV4   & 7300 &$-$  &18 :51 :14.3 & $-$06 :26 :18.9 & 15.471 & 0.799 & 0.08784   &166.000  & $\delta$ Scuti\\
 KV5   & $-$ &$-$   &18 :51 :44.0 & $-$06 :22 :01.9 & 17.685 & 1.646 & 0.06228   &165.960  & $\delta$ Scuti\\
 KV6   & 7520 &$-$  &18 :51 :18.8 & $-$06 :15 :03.1 & 15.880 & 0.835 & 0.9079   &165.930  & $\gamma$ Doradus\\
 KV7 & 4713 &$-$&18 :50 :21.6 & $-$06 :20 :40.1 & 15.048 & 0.628 & 1.06672  & 165.150  &W UMa\\
 KV8 & 4897 &$-$&18 :50 :25.8 & $-$06 :18 :58.2 & 16.484 & 0.773 & 0.554774 & 161.815  &W UMa\\
 KV9 & 4833 &$-$&18 :50 :23.7 & $-$06 :26 :18.7 & 16.641 & 1.088 & 0.64914  & 163.626  &W UMa\\
 KV10 & $-$ &$-$&18 :51 :38.8 & $-$06 :20 :56.3 & 17.051 & 1.117 & 0.51190  & 169.765  &W UMa\\
 KV11 & $-$ &$-$&18 :50 :37.6 & $-$06 :06 :45.3 & 17.501 & 1.224 & 0.55944  & 177.895  &W UMa\\
 KV12 & $-$ &$-$&18 :51 :28.1 & $-$06 :19 :14.7 & 17.734 & 1.223 & 0.490316 & 178.460  &W UMa\\
 KV13 & $-$ &$-$&18 :51 :03.1 & $-$06 :06 :58.2 & 17.883 & 0.096 & 0.3146666& 167.760  &W UMa\\
 KV14 & $-$ &$-$&18 :51 :43.5 & $-$06 :15 :17.8 & 17.970 & 1.169 & 0.41490  & 168.913  &W UMa\\
 KV15 & $-$ &$-$&18 :50 :58.9 & $-$06 :25 :08.3 & 18.345 & 0.981 & 0.388991 & 164.863  &W UMa\\
 KV16 & $-$ &$-$&18 :50 :44.5 & $-$06 :06 :06.3 & 18.543 & 0.771 & 0.3653635& 164.824  &W UMa\\
 KV17 & $-$ &$-$&18 :51 :29.8 & $-$06 :14 :06.3 & 18.649 & 1.852 & 0.5223296& 177.118  &W UMa\\
 KV18 & $-$ &$-$&18 :50 :34.1 & $-$06 :23 :35.8 & 18.670 & 1.326 & 0.43640  & 165.880  &W UMa\\
 KV19 & $-$ &$-$&18 :50 :39.0 & $-$06 :09 :56.8 & 18.703 & $-$ & 0.42900  & 164.651  &W UMa\\
 KV20 & $-$ &$-$&18 :51 :48.5 & $-$06 :19 :27.5 & 18.796 & 1.546 & 0.342800 & 165.763  &W UMa\\
 KV21 & $-$ &$-$&18 :50 :59.5 & $-$06 :06 :28.0 & 18.913 & $-$ & 0.506585 & 164.780  &W UMa\\
 KV22 & $-$ &$-$&18 :51 :26.7 & $-$06 :12 :37.3 & 19.014 & 1.530 & 0.419463 & 164.876  &W UMa\\
 KV23 & $-$ &$-$&18 :51 :35.9 & $-$06 :09 :22.7 & 19.039 & 0.924 & 0.37929  & 162.955  &W UMa\\
 KV24 & $-$ &$-$&18 :51 :04.5 & $-$06 :24 :30.4 & 19.121 & 1.150 & 0.851068 & 162.960  &W UMa\\
 KV25 & $-$ &$-$&18 :51 :48.3 & $-$06 :18 :20.6 & 19.135 & 1.708 & 0.40950  & 165.888  &W UMa\\
 KV26 & $-$ &$-$&18 :51 :15.5 & $-$06 :10 :59.0 & 19.235 & $-$ & 0.41180  & 167.895  &W UMa\\
 KV27 & $-$ &$-$&18 :51 :03.7 & $-$06 :24 :47.5 & 19.306 & $-$ & 0.46650  & 167.925  &W UMa\\
 KV28 & $-$ &$-$&18 :50 :27.4 & $-$06 :11 :58.0 & 19.353 & 1.760 & 0.32220  & 165.052  &W UMa\\
 KV29 & 1261 & 98 &18 :51 :00.1 & $-$06 :16 :37.4 & 11.918 & 0.399 & 4.64576   & 163.740  &EA\\
 KV30 & 8779 & $-$&18 :51 :47.1 & $-$06 :25 :00.6 & 15.813 & 1.414 & 1.371742  & 163.820  &EA\\
 KV31 & $-$  & $-$&18 :50 :57.3 & $-$06 :09 :40.5 & 16.824 & 1.024 & 2.67737   & 165.840  &EA\\
 KV32 & $-$  & $-$&18 :50 :20.7 & $-$06 :15 :01.7 & 17.197 &   $-$ & 0.6733    & 170.760  &EA\\
 KV33 & $-$  & $-$&18 :50 :29.6 & $-$06 :05 :39.8 & 18.023 & 0.476 & 0.65430   & 168.970  &EA\\
 KV34 & $-$  & $-$&18 :50 :41.0 & $-$06 :08 :58.8 & 18.558 & 0.863 & 1.049869  & 175.750  &EA\\
 KV35 &  965 & 99 &18 :51 :06.5 & $-$06 :14 :56.3 & 11.468 & 0.331 & $-$    & $-$  &EA\\
 KV36 &  995 &  5 &18 :51 :05.9 & $-$06 :14 :52.9 & 11.994 & 0.284 & $-$    & $-$  &EA\\
 KV37 & 1601 & 99 &18 :50 :48.7 & $-$06 :18 :15.3 & 13.647 & 0.434 & $-$    & $-$  &EA\\
 KV38 & 1731 &  0 &18 :50 :42.5 & $-$06 :16 :27.6 & 14.761 & 0.886 & $-$    & $-$  &EA\\
 KV39 &  729 & 13 &18 :51 :12.6 & $-$06 :05 :46.3 & 15.521 & 1.189 & $-$    & $-$  &EA\\
 KV40 &  230 & 67 &18 :51 :36.5 & $-$06 :07 :52.1 & 15.669 & 1.067 & $-$    & $-$  &EA\\
 KV41 & $-$  & $-$&18 :51 :44.6 & $-$06 :14 :13.4 & 16.891 & 1.086 & $-$    & $-$  &EA\\
 KV42 & $-$  & $-$&18 :50 :40.7 & $-$06 :08 :06.7 & 17.549 & 1.250 & $-$    & $-$  &EA\\
 KV43 & $-$  & $-$&18 :51 :33.6 & $-$06 :24 :30.8 & 17.717 & 1.071 & $-$    & $-$  &EA\\\hline
\end{tabular}
\tablecomments{For $\delta$ Scuti--type pulsating stars, the period and epoch are based on a dominated frequency for each variable. The third column represents the membership probability by \citet{mcnamara1977}.}
\end{table}

\clearpage
\begin{table}
\footnotesize
\caption{Results of Frequency Analysis for $\delta$ Scuti--type Stars. \label{Tab3}}
\begin{tabular}{cccccc}\hline
ID & ID$_{Hargis}$ & Frequency (c/d) &$A_{j}$\tablenotemark{\dag}(mmag) 
&$\Phi_{j}$\tablenotemark{\dag} &S/N\tablenotemark{\ddag}\\
\tableline
  HV1  &  331  & $f_1$ = 22.200 &   7.1$\pm$.8   &    3.11$\pm$.11  &    4.7 \\
       &       & $f_2$ =  9.632 &   5.4$\pm$.8   &    1.67$\pm$.14  &    3.8 \\
  HV2  &  320  & $f_1$ = 18.271 &   9.5$\pm$.5   &    4.49$\pm$.05  &    9.9 \\
  HV3  &  536  & $f_1$ = 23.758 &   6.6$\pm$1.4  &    2.74$\pm$.22  &    3.1 \\
  HV4  &  614  & $f_1$ = 21.723 &   6.1$\pm$.8   & $-$1.12$\pm$.13  &    4.6 \\
       &       & $f_2$ =  4.881 &   4.9$\pm$.8   &    2.28$\pm$.16  &    3.9 \\
  HV5  &  619  & $f_1$ = 4.179  &   8.1$\pm$1.2  & $-$0.44$\pm$.16  &    3.4 \\
       &       & $f_2$ = 27.961 &   6.0$\pm$1.2  &    1.99$\pm$.21  &    3.0 \\
  HV6  & 6870  & $f_1$ = 12.373 & 179.4$\pm$9.8  &    1.04$\pm$.06  &   10.2 \\
  KV1  &  $-$  & $f_1$ =  6.433 &  15.1$\pm$.7   &    3.33$\pm$.05  &    6.5 \\
       &       & $f_2$ = 10.417 &   6.6$\pm$.7   &    0.54$\pm$.11  &    4.6 \\
       &       & $f_3$($\approx$$f_2$$-$$f_1$$-$1) =  3.003 &   7.3$\pm$.7   & $-$0.16$\pm$.12  &    4.6 \\
       &       & $f_4$ = 12.361 &   6.6$\pm$.7   &    2.21$\pm$.11  &    4.2 \\
       &       & $f_5$ = 13.553 &   5.2$\pm$.7   &    4.35$\pm$.14  &    4.2 \\
  KV2  &  $-$  & $f_1$ = 19.388 &  89.8$\pm$2.1  &    1.20$\pm$.02  &   21.0 \\
       &       & $f_2$ = 24.835 &  19.1$\pm$2.1  & $-$0.86$\pm$.11  &    5.3 \\
  KV3  &  $-$  & $f_1$ = 21.015 &  12.7$\pm$1.6  & $-$0.74$\pm$.13  &    5.3 \\
       &       & $f_2$ = 18.435 &  10.6$\pm$1.6  &    3.04$\pm$.15  &    4.7 \\
  KV4  &  $-$  & $f_1$ = 11.384 &  13.5$\pm$1.2  &    4.64$\pm$.09  &    4.9 \\
  KV5  &  $-$  & $f_1$ = 16.056 & 116.2$\pm$4.4  & $-$0.61$\pm$.04  &   16.4 \\\hline
\end{tabular}
\tablenotetext{\dag}{$V=V_{0}+\Sigma_{j}A_{j}\rm{cos}\{2\pi f_{j}(t-t_{0})+\Phi_{j}\}$, $t_{0} = H.J.D. + 2,453,000.0$}
\tablenotetext{\ddag}{S/N = (power for each frequency / mean power after 
prewhitening for all frequencies)$^{1/2}$}
\end{table}

\end{document}